\documentclass[twocolumn,aps]{revtex4}
\input{epsfig.sty}
\newcommand{\beq}{\begin{eqnarray}}
\newcommand{\eeq}{\end{eqnarray}}
\begin{document}
\title{Sterile Neutrinos and Pulsar Velocities Revisited}
\author{Leonard S. Kisslinger\\
Department of Physics, Carnegie Mellon University, Pittsburgh, PA 15213\\ 
Mikkel B. Johnson\\
Los Alamos National Laboratory, Los Alamos, NM 87545 \\}

\begin{abstract} We calculate the momentum given to a proto neutron star
during the first 10 seconds after temperature equilibrium is reached, using 
recent evidence of sterile neutrinos and a measurement
of the mixing angle. This is a continuation of an earlier estimate with
a wide range of possible mixing angles. Using the new mixing angle we find 
that sterile neutrinos can account for the observed pulsar velocities.
\end{abstract}

\maketitle
\noindent
PACS Indices:97.60.Bw,97.60.Gb,97.60.JD
\vspace{1mm}


\section{Introduction}

  It has been observed that many pulsars move with much
greater velocities than other stars in our galaxy. This is called the
pulsar kick. See Ref.\cite {hp} for a review. Since neutrinos produced by
the URCA process dominate the emission of energy during the first 10 seconds
after the collapse of a heavy star, and for electrons in the lowest Landau
level the direction of motion is aligned
with the strong magnetic field, it was expected that active neutrino
emission would account for the pulsar velocities. It was shown, however,
that due to the small mean free path of active neutrinos this does not
work\cite{lq,al}. Using the modified URCA process, it was shown that
active neutrino emission might account for the observed pulsar velocities
during the period of about 10 to 20 seconds, when the surface of the
neutrino sphere is at the surface or the proto-neutron star\cite{hjk07}.

  A few years ago an experiment by the MiniBooNE Collaboration\cite{mini}
found that the data for electron neutrino appearance showed an excess at low 
energies, in comparison to what was expected in the standard model. 
This is consistent with the earlier LSND experiment\cite{lsnd}. An
analysis at that time\cite{ms07} indicated that there are either one
or two sterile neutrinos, but only could extract a wide range of mixing angles.
Due to the long mean path compared to active neutrinos, this suggests
that sterile neutrinos could account for the pulsar kicks. This possibility
was confirmed\cite{khj09}, but due to the uncertainly in mixing angle the
results could not be compared to observations. Prior to this pulsar kicks 
from heavier sterile neutrinos, possibly dark matter, was estimated for a 
range of parameters\cite{fkmp03}.

Recently the mixing angle in a 3+1 scenerio has been determined with much 
greater accuracy\cite{abaz12}. Last year there was a new analysis of short-
baseline neutrino oscillation data\cite{kms11} in which a 3+2 scenerio
is preferred. In the present work we will make use of the theoretical 
formulation of Ref\cite{khj09} with the new mixing angle result\cite{abaz12}, 
calculate the pulsar velocity as a function of temperature, and compare to 
observations.

\section{Asymmetric Sterile Neutrino Emissivity and Pulsar 
Velocities}

  Within about 10 seconds after the collapse of a large star about 98\% 
of neutrino emission occurs, with neutrinos produced mainly by URCA 
processes. Due to the strong magnetic field, neutrino momentum asymmetry 
is produced within the neutrinosphere if the electron is in 
the lowest Landau level, but with a small mean free path they are emitted 
only from a small surface layer of the neutrinosphere, and the pulsar kick 
cannot be accounted for. If the electron neutrino oscillates to a sterile 
neutrino, the mean free path is much greater, and pulsar velocities 
consistent with observation are possible, as found in Ref\cite{khj09}. To 
determine the pulsar velocities as a function of pulsar luminoscity or 
temperature, one must know the effective mean free path and the 
the probability, P(0), for the electron to be in the n=0 Landau level.
The expression for the asymmetric neutrino emissivity and resulting pulsar
momentum and velocity was derived in detail in Ref\cite{khj09}, and we shall
therefore only give a brief review. Then the pulsar (neutron star) velocity
will be found as a function of temperature.

\subsection{Mixing Angle in Neutrinosphere Matter}

 With the mixing angle for electron, sterile neutrinos ($\nu_e, \nu_s$)
in neutrinosphere matter =$\theta_m$, the two neutrino flavors have the form
\beq
\label{1}  
      |\nu_1> &=& cos\theta_m |\nu_e> -sin\theta_m |\nu_s> \\
      |\nu_2> &=& sin\theta_m |\nu_e> +cos\theta_m |\nu_s> \; . 
 \nonumber 
\eeq

As was shown in\cite{khj09,fkmp03}, with $\theta$ the mixing angle in vacuum, 
$sin^2(2\theta_m) \simeq sin^2(2\theta)$, therefore from Ref\cite{abaz12}
\beq
\label{2} 
         sin^2(2\theta_m) &\simeq& 0.15 \pm .05 \; .
\eeq 

Note that we used\cite{ms07} $0.004 \le sin^2(2\theta_m) \le0.2$ in 
Ref\cite{khj09}

\subsection{Neutrino Emissivity and Pulsar Momentum With a Strong 
Magetic Field}

   Using the general formulation for neutrino emissivity, see, e.g.,
Refs~\cite{fm,bw}, the main source of the asymetric emissivity that produces 
the pulsar velocity is that the electron has a large probability to 
be in the lowest (n=0) Landau level, defined as $P(0)$. A detailed derivation
of the asymetric emissivity, $\epsilon^{AS}$, is given in Ref\cite{khj09}, 
with the result
\beq
\label{4}
  \epsilon^{AS} &\simeq& 0.33 \times 10^{21} T_9^7 P(0){\rm erg\; cm^{-3}\; 
s^{-1}} \; ,
\eeq
where $T_9 = T/(10^9 K)$, with T the temperature. In Refs\cite{hjk07,khj09}
$P(0)$ was derived, with the result that $P(0) \simeq 0.4$, in agreement with  
Ref.\cite{fkmp03}. From this and Eq.(\ref{4}) one finds for the proto neutron
star momentum
\beq
\label{5}
   p_{ns} & \simeq & \frac{V_{eff}}{c} 3.3 \times 10^{21} T_9^7
 {\rm erg\; cm^{-3}} \; ,
\eeq
where $V_{eff}$ is the effective volume for the emissivity.

\subsection{Estimate of $V_{eff}$= Effective Volume for Emission}

  $V_{eff}$, the volume at the surface of the neutrinosphere from which
neutrinos are emitted, is given by the mean free path of the sterile neutrino,
$\lambda_s$, and the radius of the neutrinosphere\cite{khj09}:
\beq
\label{6}
  V_{eff} &=& (4\pi/3)( R_\nu^3-(R_\nu-\lambda_s)^3) \nonumber \\
          &\simeq& 4\pi R_\nu^2 \lambda/sin^2(2\theta) \; ,
\eeq
with $\lambda_s = \lambda/sin^2(2\theta)$, where $\lambda$ is the active
neutrino mean free path.

\subsection{Neutron Star Speed = $v_{ns}$(T)}

  Using $p_{ns}=M_{ns} v_{ns}$ and Eqs(\ref{5},\ref{6}), with the mass of the
neutron star taken as $M_{ns}=M_{sun}=2 \times 10^{33}$ gm, one 
finds with $sin^2(2\theta)$=.15
\beq
\label{7}
      v_{ns} &\simeq& 22.3 \times 10^{-7} (\frac{T}{10^{10} K})^7 \frac{km}{s} 
\; .
\eeq

During the early stages after the collapse of a massive star temperatures
T=20 MeV are expected\cite{fkmp03}. With T = 10 to 20 MeV the
pulsar velosities, with a 50\% range due to the uncertainty in $sin^2(2\theta)$,
 are shown in Fig. 1. These result can be compared to the 
pulsar velocity data as a function of luminoscity, as shown in Fig. 2.

\vspace{-2.5cm}
\begin{figure}[ht]
\epsfig{file=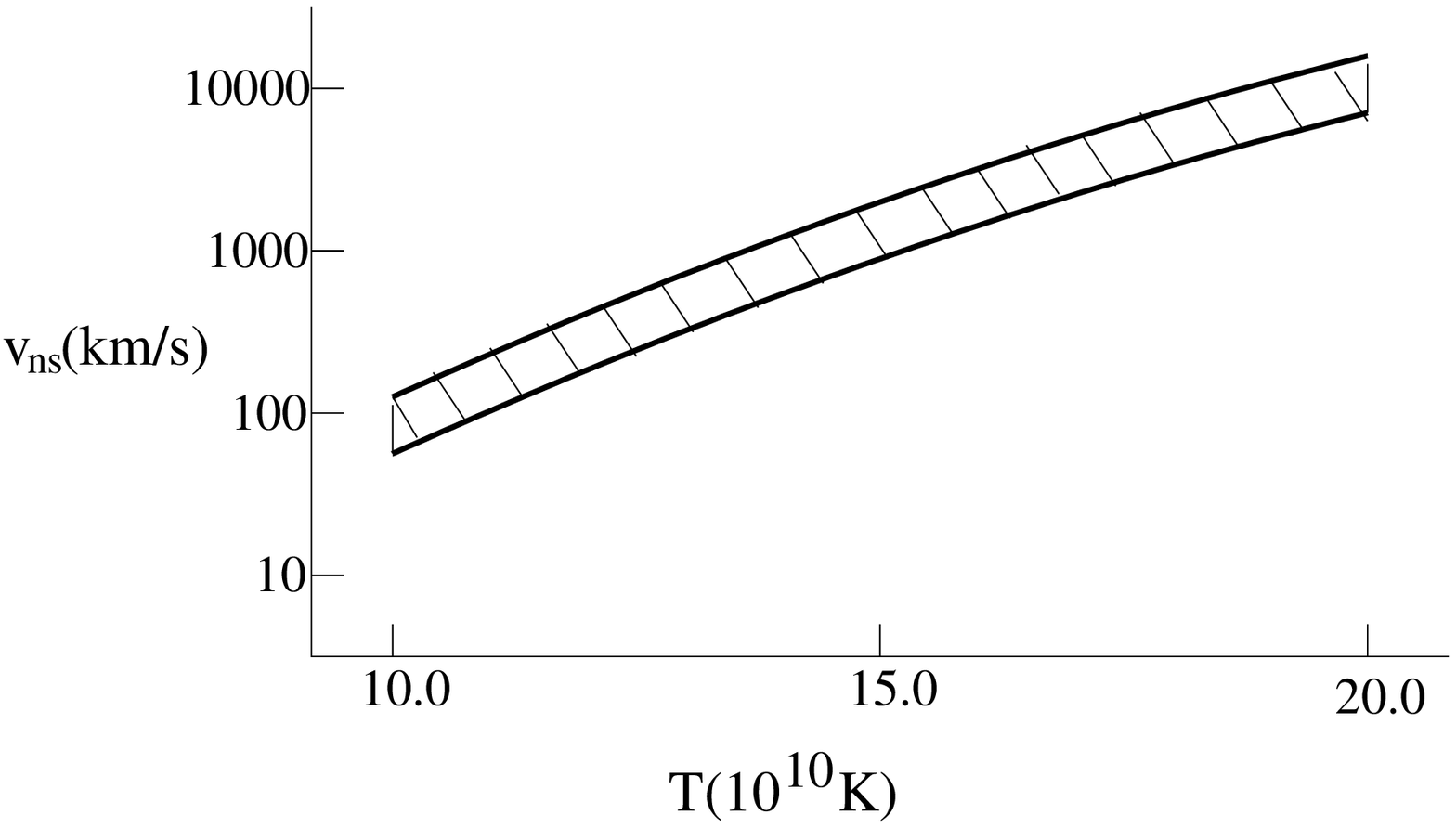,height=6cm,width=6.cm}
\caption{The pulsar velocity as a function of T for $sin^2(2\theta)=.15 \pm 
.05$}
{\label{Fig.1}}
\end{figure}
\vspace{1.5cm}

\begin{figure}[ht]
\epsfig{file=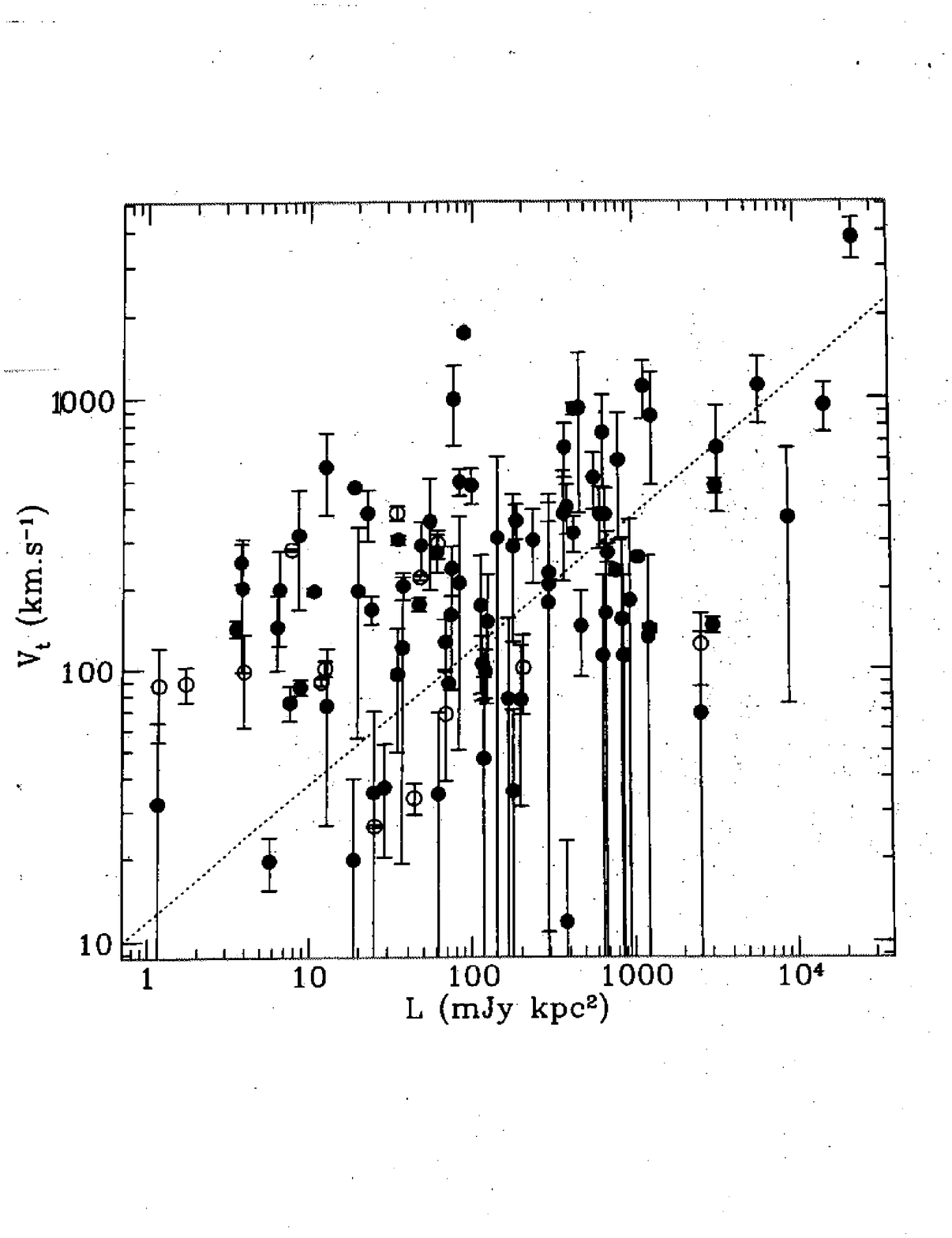,height=6cm,width=6 cm}
\caption{Pulsar velocity vs luminoscity, Ref\cite{hp97}}
{\label{Fig.2}}
\end{figure}

\newpage

\section{Conclusions}

As can be seen in Fig. 1., pulsar velocities of over 1000 km/s are predicted
from sterile neutrino emission with the mixing angle recently 
measured\cite{abaz12}.  Therefore, sterile neutrino emission can account for
the large pulsar velocities for high luminoscities (large T as in Fig. 1) 
that have been measured, as shown in Fig. 2.
This is a possible explanation of a puzzle that many have tried to explain 
for decades.
\vspace{5mm}

\Large{{\bf Acknowledgements}}\\
\normalsize
This work was supported in part by the DOE contracts W-7405-ENG-36 and 
DE-FG02-97ER41014, and in part by a grant from the Pittsburgh Foundation.
We thank Dr. William Louis for information about recent neutrino oscillation
experiments and the measurement of sterile neutrino mixing angles.

\end{document}